# THE POSSIBLE WHITE DWARF–NEUTRON STAR CONNECTION

R. CANAL AND J. GUTIERREZ
*Department of Astronomy, University of Barcelona*
*Martí i Franqués, 1 – 08028 Barcelona, Spain*

**Abstract.** The current status of the problem of whether neutron stars can form, in close binary systems, by accretion–induced collapse (AIC) of white dwarfs is examined. We find that, in principle, both initially cold C+O white dwarfs in the high–mass tail of their mass distribution in binaries and O+Ne+Mg white dwarfs can produce neutron stars. Which fractions of neutron stars in different types of binaries (or descendants from binaries) might originate from this process remains uncertain.

## 1. Introduction

Gravitational collapse of the Fe–Ni cores of massive stars (initial masses $M \gtrsim 10 - 12 \ M_{\odot}$) that have reached the end of their thermonuclear evolution is the standard mechanism to form neutron stars (NSs). A supernova explosion ejecting several solar masses of material at high velocities should simultaneously occur in order to get rid of the large mass excess of the object over the maximum possible mass of a NS ($\simeq 2.0 - 2.5 \ M_{\odot}$). The problem of transfering $\sim 1\%$ of the gravitational energy released in the collapse to the mantle and envelope of the star has not been completely solved yet, but there is little doubt that both isolated pulsars and NSs with massive companions in binaries have formed through this mechanism. NSs, however, are also found in binary systems where the companion is a low–mass ($M \sim 1 \ M_{\odot}$) star. That raised, already long ago, the question of how such systems might have survived to NS formation by the standard mechanism without being disrupted by the explosive ejection of more than half their total mass.

Canal & Schatzman (1976) (following an earlier suggestion by Schatzman 1974) first proposed that accretion of matter from the companion by a massive white dwarf (WD) in a binary might lead to the formation of a NS with little mass ejection, and presented a preliminary model in which thermonuclear explosion in the stage preceding gravitational collapse of a C+O WD was avoided. The model was further developed by Canal & Isern (1979), Canal, Isern, & Labay (1980), Isern et al. (1983), Hernanz et al. (1988), and Canal et al. (1990). Miyaji et al. (1980), Nomoto (1982, 1987), and Miyaji & Nomoto (1987) considered the case of an O+Ne+Mg WD (see Canal, Isern, & Labay 1990; Canal 1994, for reviews). This possible mechanism of nonexplosive NS formation from WDs is designated in the current literature by the term *accretion–induced collapse* (AIC) (most often without any mention to its original proponents or even with erroneous attribution to other authors: see Verbunt 1993 and reviews by van den Heuvel, for recent examples of it).

AIC of WDs has been proposed in different scenarios to explain the formation of low–mass X–ray binaries, binary pulsars with low–mass companions, binary millisecond pulsars, and single millisecond pulsars. It has even been suggested as a possible mechanism for $\gamma$–ray bursts. In all cases, however, alternative origins for the NSs involved seem also possible (such as capture of the companion by a previously formed, single neutron star). Here we will briefly review the current status of the problem of forming NSs from both C+O and O+Ne+Mg WDs.

## 2. The AIC scenario

The usual AIC *cartoon* (van den Heuvel 1981, 1984) has a WD with a mass below the Chandrasekhar mass accreting material from a low–mass companion which is filling its Roche lobe. When the Chandrasekhar mass is reached, the WD collapses to nuclear matter densities and, the mass of the newly formed object being below the maximum mass for NSs, collapse stops there. Further mass accretion produces X–ray emission and the object becomes a low–mass binary X–ray source.

If the WD were made of completely inert material (no thermonuclear burning, no electron captures), collapse would start due to general–relativistic instability when $M = 1.366\ M_\odot$, $R = 996\ km$, and $\rho_c = 2.495 \times 10^{10}\ g$, for $Z = 6$ material (Canal & Schatzman 1976). WDs, however, are actually made of materials such as He, C+O, and O+Ne+Mg, depending on at which stage of the progenitor evolution its envelope was lost. There the problems start.

— He would definitely explode at much lower densities.
— C typically ignites (explosively) at $\rho \sim (2 - 3) \times 10^9\ g\ cm^{-3}$.
— O starts to capture electrons at $\rho \sim 2 \times 10^{10}\ g\ cm^{-3}$.
— Ne does the same at $\rho \sim 9.5 \times 10^9\ g\ cm^{-3}$

— Mg begins to capture electrons at even lower densities: $\rho \sim 4 \times 10^9 \ g \ cm^{-3}$

Therefore, He WDs are absolutely excluded and in C+O WDs explosive C ignition should always precede collapse. In O+Ne+Mg WDs, $e^-$–captures (ECs) (which produce heating) also precede collapse. However, $M_{Ch} \propto Y_e^2$ ($Y_e$ being the electron mole number), and thus ECs do lower the Chandrasekhar mass. That indicates the main strategy for constructing successful AIC models: to delay explosive ignition (or reduce its effects) until ECs make $M_{Ch} < M_{WD}$.

In the *C+O case*, the preceding means keeping the material cool enough so that the $^{12}C +^{12}C$ reactions take place in the *pycnonuclear* regime (Canal & Schatzman 1976). That is not enough, however, since explosive C ignition still takes place before the start of ECs on O. Nevertheless, explosive C burning transforms the C+O mixture into matter in *nuclear statistical equilibrium* (NSE). Then ECs become fast (especially on free protons) and the Chandrasekhar mass starts decreasing. On the other hand, the explosive burning initiated at the center of the WD propagates outwards and with the energy release the star begins to expand (which would at some point cut–off the ECs). Therefore, only if the burning does not propagate very fast ($v_{burn} << c_s$, $c_s$ being the local sound velocity) we can have $M_{Ch} < M_{WD}$ before significant expansion. There, a *solid* WD, in addition to retarding C ignition, would ensure low $v_{burn}$ by suppressing hydrodynamic instabilities and allowing only the slow, purely conductive mode of burning propagation (Canal & Isern 1979).

In the *O+Ne+Mg case*, ECs would start at lower densities. They effect in heating the material and eventually inducing explosive thermonuclear burning dominates that of decreasing the Chandrasekhar mass and ECs on $^{20}$Ne trigger the explosion. Convective heat transport, however, would delay explosive ignition up to $\rho \gtrsim 2 \times 10^{10} \ g \ cm^{-3}$ and the ECs on the NSE material would then rapidly reduce $M_{Ch}$, however fast burning might propagate (Miyaji et al. 1980).

The preceding sets the stage for the current debate on the feasability of AIC from the two types of WDs: C+O and O+Ne+Mg. It involves different aspects of the physics of matter at high densities, the dynamics of thermonuclear burning, mass–accretion processes, and the evolution of intermediate–mass stars in close binary systems.

## 3. C+O white dwarfs

Nonhomologous heating upon mass accretion (the outer layers of the WD are much more compressible than the central ones) generates a "heat wave" that advances from the surface towards the center and progressively melts the solid core that might have previously formed (Hernanz et al. 1988).

Keeping to core solid until C burning starts in the pycnonuclear regime requires either very low or very high $\dot{M}$. The minimum $\dot{M}$, in the "fast" accretion case, increases with decreasing WD masses: larger initial WD masses thus favor collapse, but the maximum mass for C+O WDs is currently estimated to be $M_{C+O} \lesssim 1.1 - 1.2\ M_\odot$. Taking into account the effects of rotation on the evolution of the cores of AGB stars might, however, increase the limit almost up to the Chandrasekhar mass (Domínguez et al. 1996).

The explosive C ignition density is further sensitive to the approximations adopted for the pycnonuclear reaction rates (*static* vs *relaxed* approximations), to the crystal structure in the solid phase (blocking effects of the O nuclei on the $^{12}C + ^{12}C$ reactions), and even to the way the interpolation between the pycnonuclear and the strong screening reaction rates is done (Isern & Hernanz 1994; Isern & Canal 1994).

Even if the core melts before C ignition, burning still starts propagating conductively and collapse will occur for $\rho_{ign} \gtrsim 8.5 \times 10^9\ g\ cm^{-3}$ (García et al. 1990; Timmes & Woosley 1992). The exact ignition density depends on all of the above factors.

## 4. O+Ne+Mg white dwarfs

Those WDs should come from the evolution of stars in the $8\ M_\odot \lesssim M \lesssim 11 M_\odot$ initial mass range. The debate, here, has been mostly centered on the dependence of the explosive Ne ignition density from the treatment of the convective instability produced by EC heating in the immediately previous stages.

The very high ignition densities ($\rho_{ign} \simeq 2.5 \times 10^{10}\ g\ cm^{-3}$) found by Miyaji et al. (1980) depended on their adoption of the *Schwarzschild criterion* for the start of convection. That neglected the stabilizing effect of the $Y_e$–gradient set up by the electron captures (Mochkovitch 1984). Adoption of the *Ledoux criterion* lowered the ignition density to $\rho_{ign} \simeq (8.0 - 9.5) \times 10^9\ g\ cm^{-3}$ (Miyaji & Nomoto 1987; Canal, Isern, & Labay 1992), leaving some room to the possibility that O+Ne+Mg WDs would explode rather than collapse (Isern, Canal, & Labay 1992).

According to the most recent and accurate model calculations (Gutiérrez et al. 1996a), irrespective of the assumption made as to the development of convection, explosive ONe ignition takes place in the range $9.70 \times 10^9\ g\ cm^{-3} \leq \rho_{ign} \leq 2.12 \times 10^{10}\ g\ cm^{-3}$. That means that gravitational collapse should be the outcome in any case.

Calculations of the evolution of stars in the relevant mass range (Nomoto 1987; Domínguez, Tornambé, & Isern 1994; García–Berro & Iben 1994; Ritossa, García–Berro, & Iben 1996) have consistently shown that a fraction of C is always left unburned at the center after formation of the O+Ne (+ very little Mg) core. Upon further compression, it might ignite explosively if the C abundance were high enough. Gutiérrez et al. (1996b) have shown, however, that unless the C mass fraction were higher than about 5% (which is distinctly more than the evolutionary model predictions, with the exception of the very low–mass end of the interval) the star should not be disrupted.

## 5. Conclusions

*C+O WDs* are candidates to form NSs by AIC, provided that $\rho_{ign} \gtrsim 8.5 \times 10^9 \ g \ cm^{-3}$, but if the upper mass limit for those WDs were $M_{C+O} \lesssim 1.1 \ M_\odot$ the initial mass range might become excessively narrow. That, however, still depends on the approximations adopted for the pycnonuclear reaction rates, on the crystal structure in solid WDs, and on the way the interpolation between the pycnonuclear and the strong screening regimes is done in the models. Also, the effects of core rotation during the AGB phase might significantly increase the upper mass limit for C+O WDs.

*O+Ne(+Mg) WDs* can form NSs, unless a fraction of C larger than that predicted by current evolutionary models were left unburned at the end of the C–burning stage. NSs might also result from mergers of C+O WD pairs leading to the formation of a fast spinning O+Ne WDs which later would collapse and produce a single millisecond pulsar (Mochkovitch, Guerrero, & Segretain 1996).

There is an upper limit to the AIC rate: deleptonization of the proto–NS would produce a neutrino–driven wind carrying with it heavily neutronized material, and that should not excessively pollute the Galaxy (Woosley & Baron 1992). Work on the statistics is now still in progress, but the preliminary results indicate that AIC might explain the origin of NSs with low–mass binary companions without violating the nucleosynthesis constraints.

## References

1. Canal, R. 1994, in *Supernovae*, ed. S.A. Bludman, R. Mochkovitch & J. Zinn–Justin, Amsterdam, North–Holland, 155
2. Canal, R., García, D., Isern, J., & Labay, J. 1990, *ApJ*, **356**, L51
3. Canal, R., & Isern, J. 1979, in *White Dwarfs and Variable Degenerate Stars*, ed. H.M. Van Horn & V. Weidemann, Rochester, Univ. Rochester Press, 52
4. Canal, R., Isern, J., & Labay, J. 1980, *ApJ*, **241**, L33
5. Canal, R., Isern, J., & Labay, J. 1990, *ARA&A*, **28**, 183
6. Canal, R., & Schatzman, E. 1976, *A&A*, **46**, 229
7. Domínguez, I., Straniero, O., Tornambé, A., & Isern, J. 1996, in *Thermonuclear Supernovae*, ed. P. Ruiz–Lapuente, R. Canal, & J. Isern, Dordrecht, Kluwer, 177
8. Domínguez, I., Tornambé, A., & Isern, J. 1994, *ApJ*, **419**, 268

9. García, D., Labay, J., Canal, R., & Isern, J. 1990, in *Nuclei in the Cosmos*, ed. H. Oberhummer & W. Hillebrandt, Garching, MPA, 97
10. García–Berro, E., & Iben, I. Jr. 1994, *ApJ*, **404**, 306
11. Gutiérrez, J., García–Berro, E., Iben, I. Jr., Isern, J., Labay, J., & Canal, R. 1996a, *ApJ*, **459**, 701
12. Gutiérrez, J., Canal, R., Labay, J., Isern, J., & García–Berro, E. 1996b, in *Thermonuclear Supernovae*, ed. P. Ruiz–Lapuente, R. Canal, & J. Isern, Dordrecht, Kluwer, 303
13. Hernanz, M., Isern, J., Canal, R., Labay, J., & Mochkovitch, R. 1988, *ApJ*, **324**, 331
14. Isern, J., Canal, R., & Labay, J. 1992, *ApJ*, **372**, L83
15. Isern, J., & Canal, R. 1994, in *The Equation of State in Astrophysics*, ed. G. Chabrier & E. Schatzman, Cambridge, Cambridge Univ. Press, 186
16. Isern, J., & Hernanz, M. 1994, in *The Equation of State in Astrophysics*, ed. G. Chabrier & E. Schatzman, Cambridge, Cambridge Univ. Press, 106
17. Miyaji, S., Nomoto, K., Yokoi, K., & Sugimoto, D. 1980, *Publ. Astron. Soc. Japan*, **32**, 303
18. Miyaji, S., & Nomoto, K. 1987, *ApJ*, **318**, 307
19. Mochkovitch, R. 1984, in *Problems of Collapse and Numerical Relativity*, ed. D. Banzel & M. Signore, Dordrecht, Reidel, 125
20. Mochkovitch, R., Guerrero, J., & Segretain, L. 1996, in *Thermonuclear Supernovae*, ed. P. Ruiz–Lapuente, R. Canal, & J. Isern, Dordrecht, Kluwer, 187
21. Nomoto, K. 1982, *ApJ*, **253**, 798
22. Nomoto, K. 1987, *ApJ*, **322**, 206
23. Ritossa, C., García–Berro, E., & Iben, I. Jr. 1996, *ApJ*, in press
24. Timmes, F.S., & Woosley, S.E. 1992, *ApJ*, **396**, 649
25. van den Heuvel, E.P.J. 1981, in *Fundamental Problems in the Theory of Stellar Evolution*, ed. D. Sugimoto, D.Q. Lamb, & D.N. Schramm, Dordrecht, Reidel, 155
26. van den Heuvel, E.P.J. 1983, in *Accretion–Driven Stellar X–Ray Sources*, ed. W. H. G. Lewin & E.P.J. van den Heuvel, Cambridge, Cambridge Univ. Press, 303
27. Verbunt, F. 1993, *ARA&A*, **31**, 93
28. Woosley, S.E., & Baron, E. 1992, *ApJ*, **391**, 228